\documentclass[%
,aps%
 ,twocolumn%
 ,secnumarabic%
,amssymb, amsmath,nobibnotes, aps, prl, floatfix]{revtex4-2}
\usepackage{docs}%
\usepackage{graphicx}
\usepackage{bm}%
\usepackage[colorlinks=true,linkcolor=blue]{hyperref}%
\expandafter\ifx\csname package@font\endcsname\relax\else
 \expandafter\expandafter
 \expandafter\usepackage
 \expandafter\expandafter
 \expandafter{\csname package@font\endcsname}%
\fi

\begin{document}

\title{Defect states as a precursor of the chimera states in a ring of non-locally coupled oscillators}%
\begin{abstract}
We investigate the transition from synchronized to chimera states in a ring of non-locally coupled phase oscillators. 
Our focus is on the intermediate defect states, where solitary waves in the phase gradient profile travel at a constant speed. 
These traveling defects serve as a dynamical precursor for the nucleation of chimera clusters. 
The fraction of samples exhibiting defect states increases with the phase delay $\alpha$ and peaks at $\alpha_{c}$, where the system crosses over to asynchronous states filled with chimera clusters. 
While the traveling speed, number, and width of these defects increase with $\alpha$, 
the total spatial extent of the defects remains robust against the system size $N$.
These results shed new light on the emergence of chimera states in frustrated coupled oscillators.
\end{abstract}
\author{Tianjing Zhou}
\author{Nariya Uchida}%
\email{nariya.uchida@tohoku.ac.jp}
\affiliation{Department of Physics, Tohoku University, Sendai 980-8578, Japan}
\date{\today}%
\maketitle
\clearpage

\section{Introduction}

Chimera states, characterized by the coexistence of spatially coherent and incoherent regions, 
have become a focal point in the study of coupled oscillator 
systems~\cite{Kuramoto2002coexistence,Abrams2004chimera,Panaggio2015Nonlinearity,Omel'chenko2018The,PARASTESH20211}. In these systems, non-local coupling combined with a phase delay induces frustration and 
symmetry breaking, defying the expectation that identical oscillators with symmetric interactions 
should evolve uniformly. A variety of non-local coupling kernels—including exponential, trigonometric, 
and top-hat functions—have demonstrated the ubiquity and robustness of chimera states across 
diverse architectures. Dynamically, these states can manifest in distinct forms such as breathing~\cite{Abrams2008solvable,Laing2010Chimeras,Ma2010Robust} and traveling~\cite{Xie2014Multicluster,Dawid2019traveling,Omelchenko2020traveling} chimeras. 
Experimental evidence further confirms their existence in a wide range of platforms, from mechanical~\cite{Martens2013Chimera,Kapitaniak2014Imperfect,Carvalho2020Synchronization} 
and chemical~\cite{Tinsley2012Chimera,Wickramasinghe2013Spatially,Awal2019Smallest} oscillators to complex neural networks~\cite{Vullings2014Clustered,Bolotov2016Marginal,Kaminker2019Alternating}.

Given the large number of oscillators in some systems, such as neuronal networks, it is natural to consider 
the thermodynamic limit where the system size $N$ is much greater than the characteristic coupling range $R$. In this limit, dimensionality reduction techniques~\cite{Watanabe1994Constants,Ott2008Low} are commonly employed to map the complex dynamics onto a low-dimensional invariant manifold. This approach allows 
the derivation of partial differential equations governing the local order parameter in the continuum limit. 
Such frameworks have been instrumental in providing analytical stability analyses of chimera transitions 
and in characterizing the resulting patterns in idealized continuous systems~\cite{Laing2009The,Abrams2008solvable,Omel'chenko2013coherence,Xie2014Multicluster}.

However, discrete systems exhibit complex behavior not fully captured by continuum theories. 
Multi-headed chimera states, featuring multiple localized incoherent regions, are prominent examples 
of such complexity. 
Prior studies have analyzed the transition to these multi-headed states, exploring potential analogies 
to directed percolation~\cite{Henkel2008Non} while identifying a breakdown in universality~\cite{Duguet2019Loss,kawase2019critical,Li&Uchida2021Large-scale}. 
Within this context, "defect states" have been recognized as an intermediate phase between coherence 
and multi-headed chimera states~\cite{Duguet2019Loss}. However, the specific details of these defect states have not yet been clarified. In this paper, we revisit the transition from coherent to chimera states 
by numerically investigating defects that propagate as solitary waves in the phase gradient profile. 
By employing statistical analysis of a large ensemble, we propose a classification criterion for system states 
and examine how defect characteristics are influenced by coupling range and system size.

\section{Model}

We consider a ring of $N$ phase oscillators following the time-evolution equation
\begin{equation}
\frac{d\phi_{x}(t)}{dt} = 
\omega_{0} - \frac{1}{2R}\sum_{y, 0<|x-y|_{\rm md}\leq R}\sin(\phi_{x}(t)-\phi_{y}(t)+\alpha\pi),
\label{model}
\end{equation}
where $\phi_{x}(t)$ is the phase of the $x$-th oscillaor, $R$ is the coupling range, 
$\alpha$ is the phase delay. The intrinsic frequency $\omega_{0}$ is constant and can be set to zero 
without losing generality by the transformation $\phi_{x}-\omega_{0} t\mapsto \phi_{x}$. 
The minimum distance between two oscillators $|x-y|_{\rm md} =\min\{|x-y|, N-|x-y|\}$ is used.

We focus on the attractive coupling regime $0< \alpha<1/2$.
For quantitative description of the coherence-chimera transition, we use the local order parameter
\begin{align}
z_{\rm loc}(x,t)&=\frac{1}{2R+1}\left|\sum_{y,0\leq|x-y|_{\rm md}\leq R}e^{i\phi_{y}(t)} \right|
\label{loc}
\end{align}
and the incoherence parameter
\begin{equation}
\rho(t) = \frac{1}{N}\sum_{x=1}^{N}\left[1-z_{\rm loc}(x,t)\right], 
\label{inc}
\end{equation}
which quantifies the overall level of asynchrony in the system. 
To probe local coherence, we define the normalized phase gradient 
$\Delta \hat{\phi}_{x}(t) = \Delta \phi_x(t)/\pi$, 
where $\Delta \phi_{x}$
is the directional phase difference $\phi_{x}(t)-\phi_{x-1}(t)$
truncated into the interval $[-\pi,\pi)$. Values of $\Delta \hat{\phi}_{x} \approx 0$ correspond to 
phase-locked coherent states, whereas chimera states are characterized by a broad distribution of 
$\Delta \hat{\phi}_{x}$ across the range $[-1, 1)$. 
Detailed criteria for state classification based on $\Delta \hat{\phi}_{x}$ are provided in Sec. III.

In this paper,  coherent states refer to the $q$-twisted states~\cite{Wiley2006The}, 
\begin{equation}
    \phi_{x}(t) = \Omega t +\frac{2\pi q}{N}x+\phi_{0},
    \label{q-twisted}
\end{equation}
where $q$ is an integer and $|q|\leq(N-1)/2$.
For these states, the normalized phase gradient is constant and given by $\Delta\hat{\phi}_{x}(t) = 2q/N$.  
The corresponding local order parameter and incoherence parameter are evaluated from 
Eqs.~(\ref{loc}), (\ref{inc}) and (\ref{q-twisted}) as
\begin{align}
z_{\rm loc}(x,t)
&= \frac{1}{2R+1}
\left|1+\sum_{k=1}^{R}2\cos\left(\frac{2\pi q k}{N}\right)\right| 
\nonumber \\
&= 1-\frac{1}{2R+1}\left(\frac{2\pi q}{N}\right)^{2} \sum_{k=1}^{R} k^{2} +\mathcal{O}(q^{4})
\label{loc_q}
\end{align}
and
\begin{equation}
\rho = \frac{1}{2R+1} \left( \frac{2\pi q}{N} \right)^2 \sum_{k=1}^{R} k^2 + \mathcal{O}(q^4),
\label{inc_q}
\end{equation}
respectively. The latter provides a theoretical baseline for distinguishing twisted states 
in our numerical results.

\section{Numerical Results}

We numerically integrated Eq.~(\ref{model}) using the fourth-order Runge-Kutta method with 
a time step $\Delta t = 0.01$. For each sample, the initial state was a spatially random phase configuration 
with phases chosen uniformly from $[0, 2\pi)$. 
Unless otherwise stated, we used a system size of $N = 1000$ and a coupling range of $R = 5$, 
and systematically varied the phase lag $\alpha$.

\begin{figure*}
\centering
\includegraphics[width=0.8\linewidth]{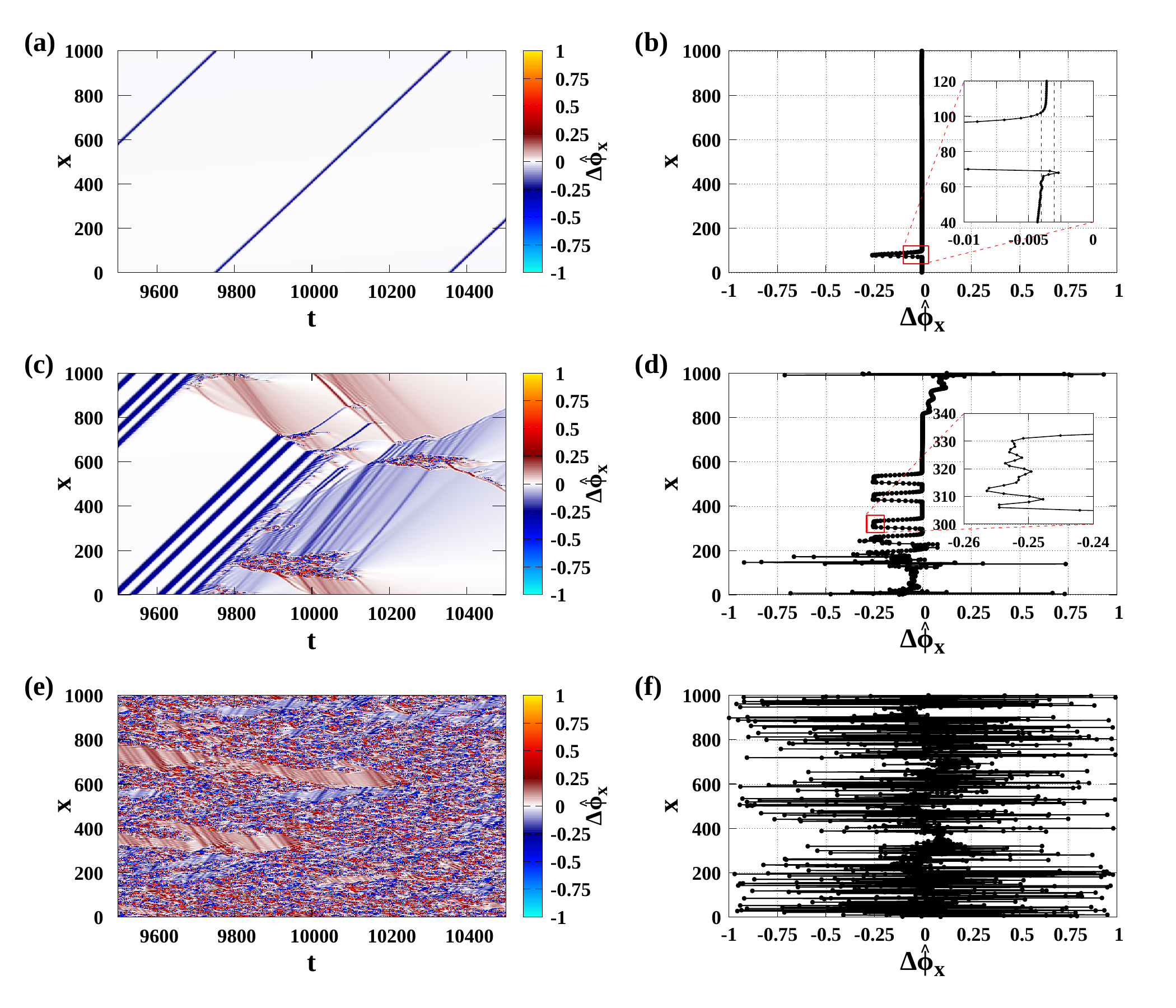}
\caption{Spatiotemporal maps (left column) and spatial profiles of $\Delta\hat{\phi}_{x}$ at $t=9800$ 
(right column) for $N=1000$ and $R=5$. 
(a), (b) $\alpha=0.4$: defect states. 
(c), (d) $\alpha=0.437$: mixture of defect states and chimera states. 
(e), (f) $\alpha=0.45$: chimera states. 
The inset of (b) and (d) shows magnified plots with dotted lines in (b) 
corresponding to $\Delta \hat{\phi}_{x} =-0.004, -0.003$.
}
\label{fig1}
\makeatletter    \def\@currentlabel{\thefigure(a)}\label{fig1_a}
\def\@currentlabel{\thefigure(b)}\label{fig1_b}
\def\@currentlabel{\thefigure(c)}\label{fig1_c}
\def\@currentlabel{\thefigure(d)}\label{fig1_d}
\def\@currentlabel{\thefigure(e)}\label{fig1_e}
\def\@currentlabel{\thefigure(f)}\label{fig1_f}
\makeatother
\end{figure*}

Figure 1 shows the spatiotemporal evolution and spatial profiles of the phase gradient $\Delta\hat{\phi}_x$ for different values of $\alpha$. 
At $\alpha = 0.4$ [Figs.~\ref{fig1_a} and \ref{fig1_b}], we observe a solitary wave that propagates at a constant speed. 
Following Ref.~\cite{Duguet2019Loss}, we refer to this structure as a defect and the corresponding 
collective phase as the defect state. While such defects were previously interpreted as interfaces between 
different $q$-twisted states, our results reveal a more subtle structure. 
As shown in the inset of Fig.~\ref{fig1_b}, the phase gradient in the coherent regions on either side of a defect 
does not necessarily match the discrete values expected for integer $q$. This deviation suggests that 
the regions surrounding a defect are not perfectly synchronized twisted states.

As $\alpha$ increases to 0.437 [Figs.~\ref{fig1_c} and \ref{fig1_d}], multiple defects appear, and chimera clusters characterized by large phase-gradient fluctuations begin to emerge from them. These clusters are clearly distinguishable from the traveling defects, which maintain a maximum phase gradient of $|\Delta\hat{\phi}_{x}| \sim 0.25$ [Fig.~\ref{fig1_d}, inset]. 

At $\alpha = 0.45$ [Figs.~\ref{fig1_e} and \ref{fig1_f}], the system reaches a multi-headed chimera state. 
In the spatiotemporal map [Fig.~\ref{fig1_e}], the slanted structures that maintain a constant slope correspond to 
persistent defects. These defects coexist with incoherent regions, which form the multi-headed chimera appearing as broad, irregular strips separated by narrow coherent intervals. 
The corresponding spatial profile [Fig.~\ref{fig1_f}] confirms that these regions are characterized by large-amplitude, random phase-gradient fluctuations spanning the entire range $[-1, 1)$. This regime represents a state where the regular defect pattern is largely replaced by the proliferation of incoherent chimera clusters.

These observations suggest that the regular defect state serves as a dynamical precursor to the chimera state. However, due to the sensitive dependence on initial conditions, this intermediate phase is not captured in every sample. Relying on a single initial configuration may therefore result in missing the defect state entirely. This motivates us to perform a large-ensemble statistical analysis to characterize the collective behavior of the system under random initial phases. We also note that, while our figures focus on negative phase gradients, 
we have observed defects with positive $\Delta \hat{\phi}_{x}$ values propagating 
in the opposite ($-x$) direction, reflecting the underlying symmetry of the system.

\begin{figure*}
\centering
\includegraphics[width=\linewidth]{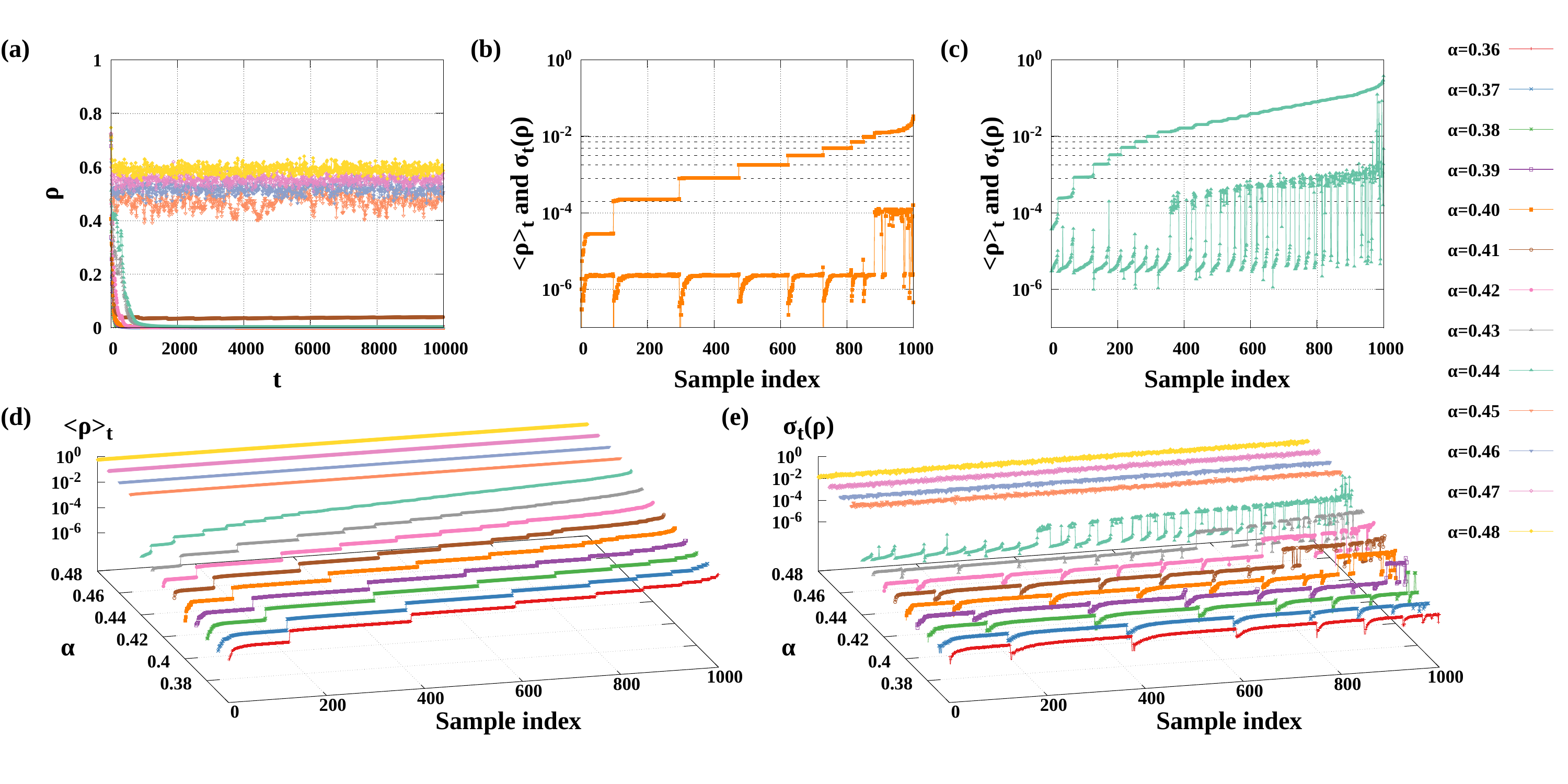}
\caption{Time evolution and time-dependent statistics of the incoherence parameter $\rho$. 
For $N=1000$ and $R=5$: (a) time evolution of $\rho$ for 13 different samples 
with $\alpha$ ranging from 0.36 to 0.48; (b), (c), (d), (e) time average and 
time standard deviation of the incoherence parameter, $\langle\rho\rangle_{t}$ 
and $\sigma_t(\rho)$, for 1000 samples with different random initial configurations. 
The statistics are taken within the time window $t=9000\text{--}10000$. 
Panels (b) and (c) show results for $\alpha = 0.40$ and $0.44$, respectively; 
the dotted lines 
indicate the theoretical values of $\rho$ for $q$-twisted states with $q=1\text{--}7$.
}
\label{fig2}
\makeatletter    \def\@currentlabel{\thefigure(a)}\label{fig2_a}
\def\@currentlabel{\thefigure(b)}\label{fig2_b}
\def\@currentlabel{\thefigure(c)}\label{fig2_c}
\def\@currentlabel{\thefigure(d)}\label{fig2_d}
\def\@currentlabel{\thefigure(e)}\label{fig2_e}
\makeatother
\end{figure*}

Fig.~\ref{fig2_a} illustrates the time evolution of the incoherence parameter $\rho$ for various 
initial conditions and $\alpha$ values. Since the system reaches a steady state after 
approximately $t \approx 1000$, we perform the subsequent statistical analysis within 
the window $t \in [9000, 10000]$. Figs.~\ref{fig2_b} and \ref{fig2_c} show the 
time-averaged incoherence parameter $\langle\rho\rangle_t$ and its standard deviation $\sigma_t(\rho)$ 
for 1000 samples at $\alpha = 0.40$ and $0.44$, respectively, ordered by $\langle\rho\rangle_t$. 
The stair-shaped plateaus in $\langle\rho\rangle_t$ reflect the multistability of $q$-twisted states. 
As indicated by the dotted lines, seven of the steps align precisely with the theoretical values for 
$|q| = 1 \text{--} 7$ derived from Eq.~(\ref{inc_q}). These coherent states are characterized by a vanishingly small 
$\sigma_t(\rho)$, as all oscillators share a common phase velocity. Notably, the standard deviation 
$\sigma_t(\rho)$ exhibits a distinct three-layer hierarchy that serves as a robust criterion for state classification:
(i) Coherent layer: The bottom layer ($\sigma_t(\rho) \ll 10^{-4}$) corresponds to 
the $q$-twisted states. 
(ii) Defect layer: The intermediate layer represents the defect states, where the propagation of solitary waves
introduces moderate, persistent fluctuations. 
(iii) Chimera layer: The top layer corresponds to chimera states characterized by large-amplitude fluctuations.
As $\alpha$ increases [Figs.~\ref{fig2_d} and \ref{fig2_e}], the fraction of samples residing in the coherent steps systematically decreases. This step structure collapses at $\alpha = 0.45$, indicating the disappearance of purely twisted states in our ensemble. Conversely, the population of defect states increases for 
$\alpha \leq 0.44$ before being replaced by the proliferation of chimera clusters at $\alpha \geq 0.45$.

\begin{figure}[h]
\centering
\includegraphics[width=0.85\linewidth]{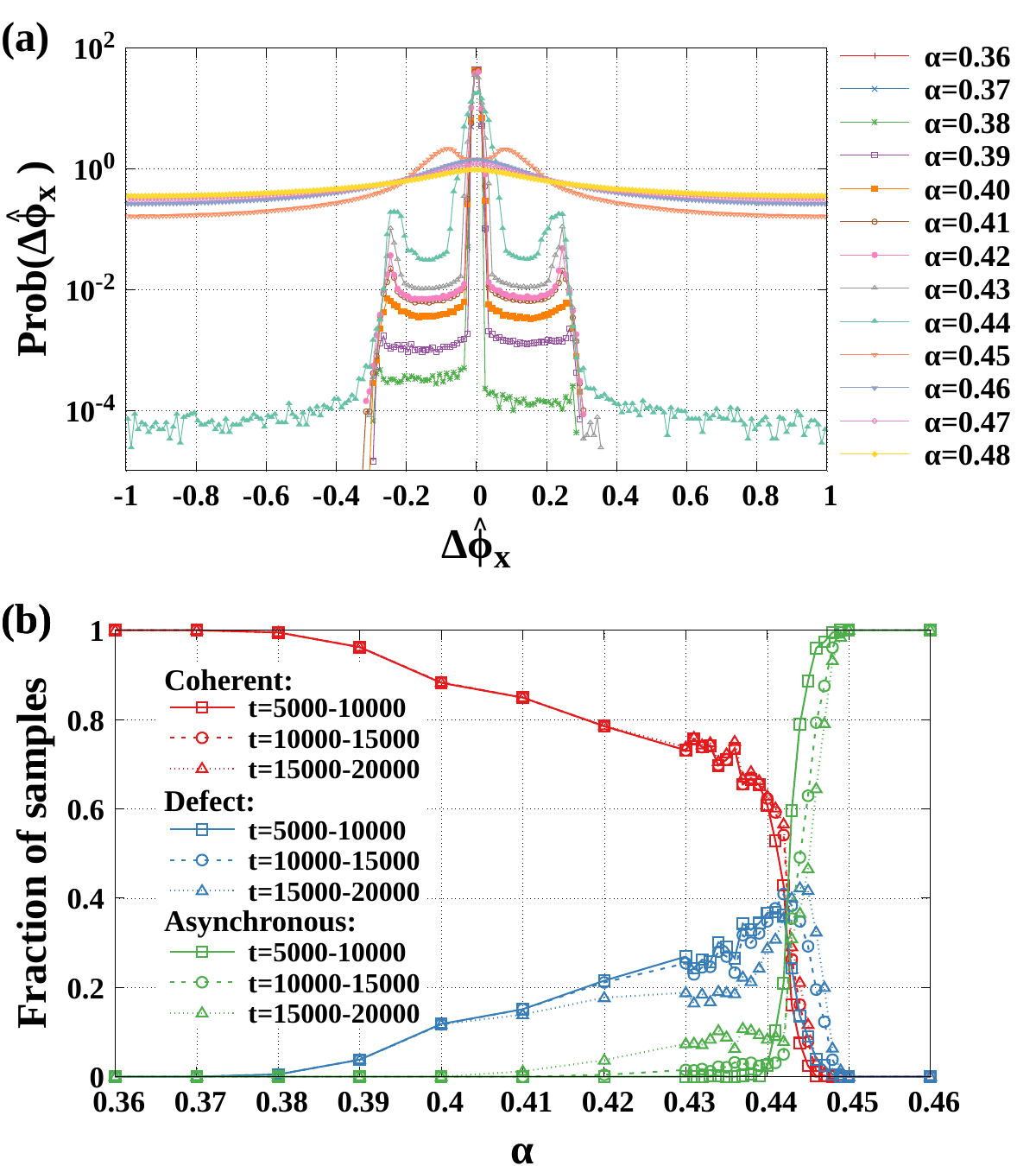}
\caption{Statistical characteristics of defects and state transitions for $R=5$ and $N=1000$. 
(a) Normalized distribution $\text{Prob}(\Delta \hat{\phi}_x)$ of the phase gradient $\Delta \hat{\phi}_{x}$,
computed from 1000 samples for each $\alpha$ within the time window $t = 9000\text{--}10000$. 
(b) Fraction of samples in three different system states: coherent (red), defect (blue), 
and asynchronous (green). The results are shown for three consecutive time intervals: 
$t = 5000\text{--}10000$ (early stage, solid lines), 
$10000\text{--}15000$ (middle stage, dashed lines), and 
$15000\text{--}20000$ (late stage, dotted lines).
}
\label{fig3}
\makeatletter    \def\@currentlabel{\thefigure(a)}\label{fig3_a}
\def\@currentlabel{\thefigure(b)}\label{fig3_b}
\makeatother
\end{figure}

To further quantify the emergence of defects, we examine the statistical distribution of the phase gradient. Fig.~\ref{fig3_a} shows the normalized distribution $\text{Prob}(\Delta \hat{\phi}_x)$ for various values of 
$\alpha$. For small $\alpha$, the distribution is sharply peaked at $\Delta \hat{\phi}_x = 0$, reflecting the dominance of coherent states. As $\alpha$ increases, two satellite peaks emerge near $|\Delta \hat{\phi}_x| \approx 0.25$. These symmetric peaks correspond to the characteristic phase gradient of the traveling defects 
propagating in either the $+x$ or $-x$ direction, as observed in Figs.~\ref{fig1_b} and \ref{fig1_d}. 
In the asynchronous regime ($\alpha \geq 0.44$), the distribution becomes significantly broader and 
the satellite peaks are subsumed by a wide pedestal, indicating the high spatial irregularity 
of the multi-headed chimera states. 

Based on the spatial features of the phase gradient $|\Delta\hat{\phi}_{x}(t)|$ within 
a given time window, we categorize the collective states into three groups:
\begin{itemize}
\item Coherent states: $\max_{x,t} |\Delta\hat{\phi}_{x}(t)| \leq 0.1$.
\item Defect states: $0.1<\max_{x,t}|\Delta\hat{\phi}_{x}(t)| < 0.5$.
\item Asynchronous states: $\max_{x,t} |\Delta\hat{\phi}_{x}(t)| \geq 0.5$.
\end{itemize}
Samples containing both defects and chimera regions are classified as asynchronous, as the high-amplitude phase fluctuations represent the fundamental asynchrony of the system. 
Fig.~\ref{fig3_b} shows the fraction of 1000 random-initiated samples for each state as a function of $\alpha$. 
At low $\alpha$ (0.36–0.37), the system is exclusively coherent. Defect states emerge at $\alpha \approx 0.38$ and their fraction increases steadily, reaching a peak near $\alpha = 0.44$. 
This peak coincides with the sharp decline of the coherent fraction and the rapid onset of asynchronous states,
confirming that the defect state serves as an intermediate phase in the transition to multi-headed chimeras.

Furthermore, we examined the stability of these fractions across three different time windows. 
While the results for the first two windows ($t = 5000\text{--}10000$ and $10000\text{--}15000$) 
are nearly identical, the latest window ($t = 15000\text{--}20000$) reveals a subtle yet noticeable shift
specifically in the range $\alpha \geq 0.41$: 
the defect fraction decreases while the asynchronous fraction increases. 
This temporal trend suggests that the defect state may represent a long-lived transient within a relaxation
process. Although the present data do not allow for a definitive conclusion regarding the long-time limit, 
the gradual nucleation of asynchronous regions from defects points toward the possibility that the 
defect-dominated regime is influenced by finite-time effects, especially near the coherent-asynchronous
crossover.
\begin{figure}[h]
\centering
\includegraphics[width=0.8\linewidth]{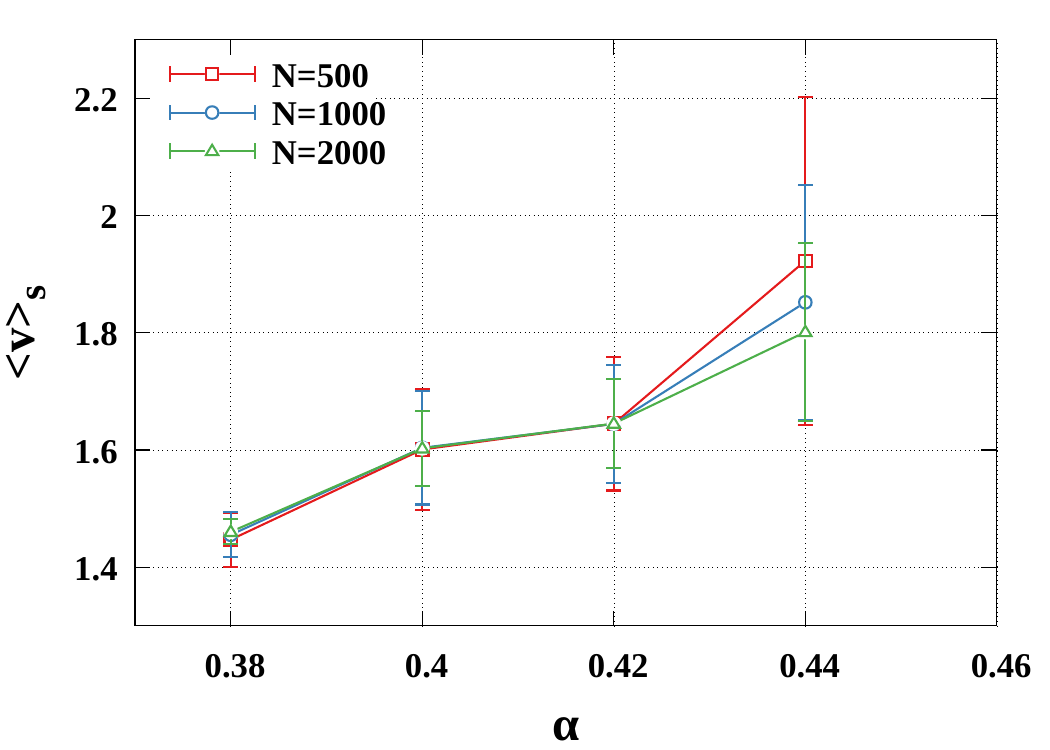}
\caption{Sample average traveling speed $\langle v \rangle_s$ of defects as a function of $\alpha$ for $R=5$.
The statistics are calculated from all defect-state samples identified within an ensemble of 1000 samples for each $N$ and $\alpha$. The average is taken over the time window $t = 9000\text{--}10000$. 
Results for three different system sizes are shown: $N=500$ (red squares), $N=1000$ (blue circles), 
and $N=2000$ (green triangles).
}
\label{fig4}
\end{figure}

Fig.~\ref{fig4} shows the sample average traveling speed $\langle v \rangle_s$ of the defects. 
To calculate $v$ for an individual sample, we track the spatial position $x^{\ast}(t)$ of the defect, 
defined as the smallest index $x$ where $|\Delta \hat{\phi}_x(t)|$ crosses the threshold of 0.1. 
Within a specified time window $t_0\text{--}t_1$, the speed is determined 
by $v = |x^{\ast}(t_1) - x^{\ast}(t_0)| / (t_1 - t_0)$. The sample average $\langle v \rangle_s$ is then computed across all samples classified as defect states.

As $\alpha$ increases, both the average traveling speed $\langle v \rangle_s$ and its standard deviation 
(error bars) increase. This trend suggests that higher propagation speeds and wider statistical dispersion 
are hallmarks of the transition from stable defect states to the asynchronous regime.

Notably, for $0.38 \le \alpha \le 0.42$, the values of $\langle v \rangle_s$ for all system sizes $N$ coincide,
confirming that the defect dynamics are independent of $N$ in this range. 
However, this overlap breaks down at $\alpha = 0.44$. This deviation occurs precisely at the 
same $\alpha$ value where asynchronous (chimera) states first emerge in the ensemble [Fig.~\ref{fig3_b}], 
suggesting that the onset of chimera regions begins to perturb the intrinsic propagation of the defects.

\begin{figure*}
\centering
\includegraphics[width=0.85\linewidth]{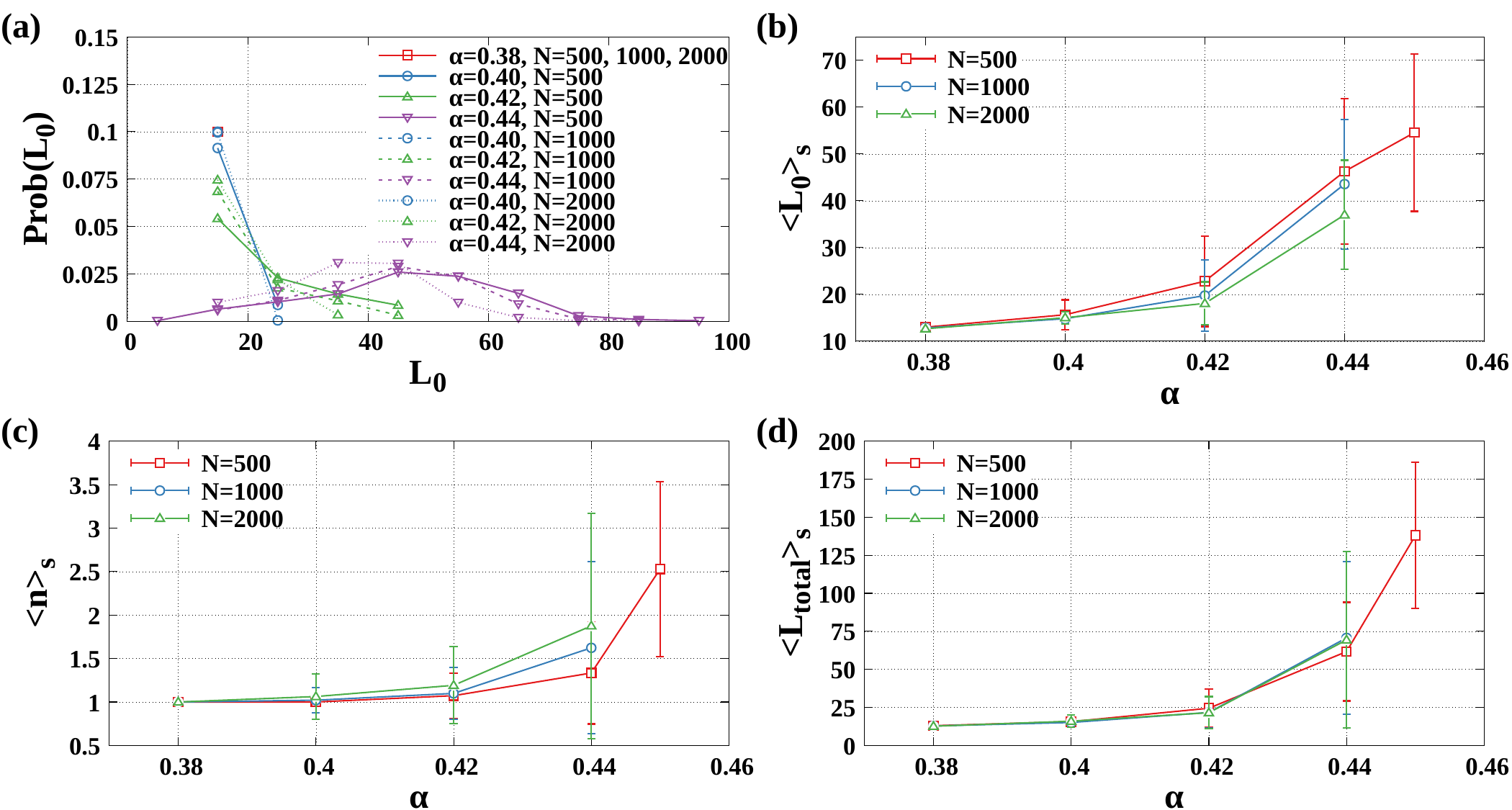}
\caption{Statistical characteristics of defect width and number for $R=5$. 
(a) Normalized distribution $\text{Prob}(L_0)$ of the individual defect width $L_0$. 
(b) Sample average of the individual defect width $\langle L_0 \rangle_s$, 
(c) sample average number of defects $\langle n \rangle_s$, and 
(d) sample average of the total defect width $\langle L_{\rm total} \rangle_s$, 
plotted as functions of $\alpha$. Statistics are computed from all samples identified 
as defect states within an ensemble of 1000 samples for each $N$ and $\alpha$.
}
\label{fig5}
\makeatletter    \def\@currentlabel{\thefigure(a)}\label{fig5_a}
\def\@currentlabel{\thefigure(b)}\label{fig5_b}
\def\@currentlabel{\thefigure(c)}\label{fig5_c}
\def\@currentlabel{\thefigure(d)}\label{fig5_d}
\makeatother
\end{figure*}

We have examined the statistical properties of the number and width of defects in detail, as shown in Fig.~\ref{fig5}. 
To define the width of a single defect $L_0$ at a fixed time $t = 10000$, we identify the spatial regions 
where the phase gradient exceeds the threshold $|\Delta \hat{\phi}_x| > 0.1$. The defect width is then determined by the length of these regions along the $x$-coordinate.

Fig.~\ref{fig5_a} presents the normalized distribution $\text{Prob}(L_0)$ for various values of $\alpha$ and $N$. For the range $\alpha = 0.38\text{--}0.42$, the distribution is dominated by narrow defects with widths 
$L_0 < 20$. In contrast, at $\alpha = 0.44$, the distribution becomes noticeably flatter, with a central peak 
shifting to approximately $40\text{--}50$. This shift indicates that the defect width distribution undergoes 
a qualitative change as the system approaches the chimera transition. Furthermore, a slight size effect is observed, where larger systems tend to exhibit narrower defects more frequently.

Figs.~\ref{fig5_b} and \ref{fig5_c} show the sample average of the individual defect width $\langle L_0 \rangle_s$ and the average number of defects $\langle n \rangle_s$, respectively. The average width $\langle L_0 \rangle_s$ is positively correlated with $\alpha$ but shows a negative correlation with $N$. Conversely, the average number of defects $\langle n \rangle_s$ is positively correlated with both $\alpha$ and $N$. Finally, it is worth noting that in Fig.~\ref{fig5_d}, the curves for the sample average total defect width $\langle L_{\rm total} \rangle_s$ coincide quite well for $N=500\text{--}2000$ over a wide range of $\alpha$. 

\begin{figure*}
\centering
\includegraphics[width=1\linewidth]{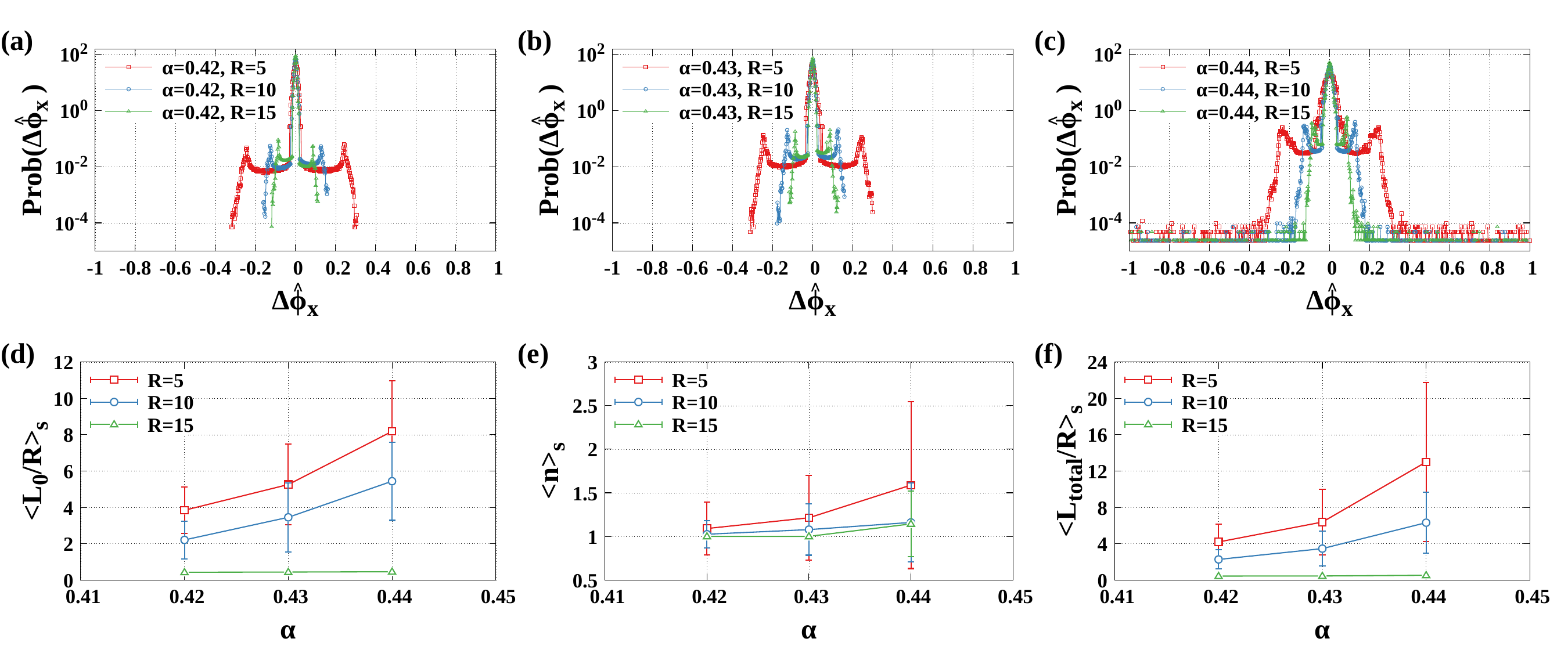}
\caption{Statistical characteristics of defects with respect to the coupling range $R$ for $N=1000$. 
(a)–(c) Normalized distribution $\text{Prob}(\Delta \hat{\phi})$ of the phase gradient for 
(a) $\alpha = 0.42$, (b) $\alpha = 0.43$, and (c) $\alpha = 0.44$. 
These statistics are taken from all 1000 samples for each $\alpha$ and $R$ within the time 
window $t = 9000\text{--}10000$. 
(d) Sample average of the relative defect width $\langle L_0/R \rangle_s$, 
(e) sample average number of defects $\langle n \rangle_s$, and 
(f) sample average of the relative total defect width $\langle L_{\rm total}/R \rangle_s$, 
plotted as functions of $\alpha$. 
The statistics in (d)–(f) are computed specifically from samples identified as defect states within 
the ensemble of 1000 samples for each $\alpha$ and $R$.
}
\label{fig6}
\makeatletter 
\def\@currentlabel{\thefigure(a)}\label{fig6_a}
\def\@currentlabel{\thefigure(b)}\label{fig6_b}
\def\@currentlabel{\thefigure(c)}\label{fig6_c}
\def\@currentlabel{(c)}\label{fig6_cc}
\def\@currentlabel{\thefigure(d)}\label{fig6_d}
\def\@currentlabel{\thefigure(e)}\label{fig6_e}
\def\@currentlabel{\thefigure(f)}\label{fig6_f}
\makeatother
\end{figure*}

We further investigate the dependency of defect characteristics on the coupling range $R$ for $N=1000$
systems, as displayed in Fig.~\ref{fig6}. Figs.~\ref{fig6_a}--\ref{fig6_cc} show the normalized distribution 
$\text{Prob}(\Delta \hat{\phi}_x)$ for different values of $\alpha$. 
For all tested coupling ranges ($R=5, 10$, and $15$), asynchronous states emerge at $\alpha \approx 0.44$, 
suggesting a common transition threshold across these scales. However, we observe that the positions of 
the satellite peaks in $\Delta \hat{\phi}_x$ are inversely correlated with $R$. This shift indicates that 
the characteristic phase gradient of a defect decreases as the coupling range expands.

The spatial scale of the defects also exhibits a complex dependence on $R$. While the absolute average width $\langle L_0 \rangle_s$ (not shown) exhibits a non-monotonic dependence on $R$, we find that the relative width $\langle L_0/R \rangle_s$ reveals a clearer trend. As shown in Fig.~\ref{fig6_d}, by introducing $R$ as a characteristic length scale, $\langle L_0/R \rangle_s$ is found to decrease monotonically as $R$ increases.

Figs.~\ref{fig6_e} and \ref{fig6_f} present the average number of defects $\langle n \rangle_s$ and the relative total defect width $\langle L_{\rm total}/R \rangle_s$, respectively. While $\langle n \rangle_s$ for $R=10$ and $15$ show similar values, an overall negative correlation with $R$ is evident. Similarly, the relative total width $\langle L_{\rm total}/R \rangle_s$ decreases as $R$ increases. In summary, we observe a reduction in defect height (phase gradient), relative width $L_0/R$, and total spatial extent $L_{\rm total}/R$ with increasing $R$. These results indicate that the stability and prevalence of these localized structures are significantly influenced by the coupling range $R$.

\section{Discussion and summary}\label{sec:front}

We found that perturbations within the defect states amplify and eventually develop into chimera states. 
Within the investigated parameter region ($N=1000\text{--}2000$, $R=5\text{--}15$), 
defect states appear across a wide range of phase-lag $\alpha$ as a precursor to the emergence of 
chimera clusters. These intermediate states can be best identified by the emergence of symmetric 
satellite peaks in the phase-gradient distribution, and thus we adopted $\Delta \hat{\phi}_{x}$ as 
a more sensitive probe than the incoherence parameter $\rho$ to distinguish the various collective states. Furthermore, the observed decrease in the defect fraction over long time scales [Fig.~\ref{fig3_b}] suggests that these precursors may represent long-lived transients within a slow relaxation process toward the fully developed asynchronous regime, particularly near the crossover point.

The accumulation of phase differences that are incompatible with integer $q$-values within these defects is a characteristic feature of one-dimensional systems, where $q$-twisted states constitute a fundamental part of the coherent phase space. In two-dimensional systems, plane waves are typically not observed as stable patterns under attractive coupling. Consequently, the transition from synchronized to chimera states in higher dimensions likely follows a distinct dynamical trajectory, differing from the defect-mediated path observed here.

While the defect width $L_0$ was previously suggested to scale as $\mathcal{O}(R)$ , our results show that $L_0/R$ actually decreases with $R$. It is therefore important to clarify whether these defects disappear in the limit $R \to \infty$ while keeping the ratio $N/R$ constant and large ($N/R \gg 1$). Such an investigation would require even larger system sizes and analysis in the continuous limit. 
Additionally, the total defect width $\langle L_{\rm total} \rangle_{\rm s}$ appears consistent across $N=500\text{--}2000$, 
whereas it exhibits a distinct dependence on the coupling range $R$. This contrast suggests that the spatial extent of the precursor phase is governed by the local interaction scale rather than the global system size, reflecting a robust physical characteristic of the defect-mediated nucleation.

The role of defect states in the breakdown of universality has been previously noted; specifically, the critical behavior at the onset of asynchronous states was found to differ from the 1-D directed percolation (DP) universality class due to these defects~\cite{Duguet2019Loss}. Our results provide a more detailed physical basis for this departure by revealing the $R$-dependent statistics of the defect states. Because the defects act as a precursor to the chimera transition and their properties vary with the coupling range, the resulting transition naturally deviates from the standard DP class. 

Our work provides a point of contrast with previous research on non-uniform twisted states. Modulated twisted states were discussed in systems of oscillators with Lorentzian intrinsic frequency distributions~\cite{Xie2019twisted}.
Non-uniformly twisted states were also attained 
in systems using trigonometric coupling kernels 
with the lowest-order spatial Fourier modes~\cite{Smirnov2024Nonuniformly}. 
In contrast, our research employs a top-hat coupling kernel for identical oscillators. 
The absence of frequency dispersion allows the transition to be captured as a sharp crossover, characterized by the distinct peaking of defect states.
The kernel naturally incorporates high-order spatial Fourier modes, which are essential for observing the multi-headed chimera behavior and the specific defect dynamics reported here.

\bibliographystyle{apsrev4-1}
\bibliography{zhou2026defect}

\end{document}